\documentclass[9pt]{extarticle}
\usepackage{spconf,amsmath,graphicx}
\usepackage{amssymb}
\usepackage{caption}
\usepackage{subcaption}
\usepackage{booktabs}
\usepackage{multirow}
\usepackage{adjustbox}
\usepackage{makecell} 
\usepackage[table]{xcolor} 
\usepackage{hyperref}
\usepackage{url}

\title{PhoenixCodec: Taming Neural Speech Coding for Extreme Low-Resource Scenarios}

\name{\begin{minipage}{\textwidth}\centering
      Zixiang Wan$^{\star\dagger}$\thanks{$^{\dagger}$Equal contribution.} \thanks{$^{*}$Corresponding authors.} 
      \thanks{Project Page: \url{https://ggiggit.github.io/phoenixcodec.github.io/}}
      \quad Haoran Zhao$^{\ddagger\dagger}$\footnotemark[1]
      \quad Guochang Zhang$^{\ddagger}$
      \quad Runqiang Han$^{\ddagger}$ \\
      Jianqiang Wei$^{\ddagger*}$
      \quad Yuexian Zou$^{\star*}$\footnotemark[2]
      \end{minipage}}

\address{\begin{minipage}{\textwidth}\centering
         $^{\ddagger}$ Audio Innovation Technology Department, Anker Inc, Beijing, China \\
         $^{\star}$ Guangdong Provincial Key Laboratory of Ultra High Definition Immersive Media Technology, \\
         Peking University, Shenzhen, China \\
         \href{mailto:zxwan25@stu.pku.edu.cn}{\texttt{zxwan25@stu.pku.edu.cn}}, 
         \href{mailto:alex.wei@anker-in.com}{\texttt{alex.wei@anker-in.com}}
         \end{minipage}}
         
\begin{document}
%
\maketitle
\begin{abstract}
This paper presents PhoenixCodec, a comprehensive neural speech coding and decoding framework designed for extremely low-resource conditions. The proposed system integrates an optimized asymmetric frequency–time architecture, a Cyclical Calibration and Refinement (CCR) training strategy, and a noise-invariant fine-tuning procedure. Under stringent constraints—computation below 700 MFLOPs, latency less than 30 ms, and dual-rate support at~1~kbps and~6~kbps—existing methods face a trade-off between efficiency and quality. PhoenixCodec addresses these challenges by alleviating the resource-scattering of conventional decoders, employing CCR to enhance optimization stability, and enhancing robustness through noisy-sample fine-tuning. In the \textit{LRAC~2025 Challenge Track~1}, the proposed system ranked third overall and demonstrated the best performance at 1 kbps in both \textit{real-world noise and reverberation} and \textit{intelligibility in clean} tests, confirming its effectiveness.

\end{abstract}
\section{Introduction}
\label{sec:intro}

With the increasing prevalence of online meetings, instant messaging, and other forms of real-time speech communication, the demand for transmitting high-quality audio over low-bandwidth networks has grown rapidly. 
Neural speech codecs have emerged as a promising solution, delivering superior performance at low bitrates compared to conventional approaches~\cite{yang2025ucodecultralowframerate, chen2025auvteachingaudiouniversal, yang2025almtokenizer}.

Recent advances in neural speech codecs have achieved significant improvements in reconstruction quality~\cite{jiang2025unicodec, ye2025codec, welker2025flowdec}.
Representative models such as SoundStream \cite{zeghidour2021soundstream}, EnCodec \cite{defossez2022high}, and DAC \cite{kumar2023high} employ optimized symmetric time-domain architectures, approaching transparent speech reproduction even at low bitrates.
Building on these advancements, subsequent research has focused on codec design under diverse resource constraints: LSCodec \cite{guo2024lscodec} and BigCodec \cite{xin2024bigcodec} enhance performance in the ultra-low-bitrate regime through information disentanglement and capacity scaling, while FreqCodec \cite{du2024funcodec} and SpecTokenizer \cite{wan2025spectokenizer} exploit efficient frequency-domain designs to reduce computational cost, and Lyra-V2 \cite{Google2022LyraV2} delivers production-ready, low-latency streaming capability.

Although remarkable progress has been made, achieving an optimal balance under the \textit{LRAC~2025}\footnote{\url{https://crowdsourcing.cisco.com/lrac-challenge/2025/}} \cite{wojcicki2025lowresourceaudiocodeclrac} constraints—dual-rate support (1~and~6~kbps), limited complexity ($\leq$~700~MFLOPs), and real-time latency ($\leq$~30~ms)—remains highly challenging.
This calls for a codec that unifies representational efficiency, reconstruction quality, and hardware adaptability within a single design.

To advance this frontier, we introduce \textbf{PhoenixCodec}, a comprehensive system for extreme low-resource conditions integrating innovations in architecture, training, and fine-tuning methodology.

Our key contributions are summarized as follows:

\begin{itemize}
  \item \textbf{Frequency–Time Domain Asymmetric Architecture.}  
  We propose and validate an efficient hybrid architecture that combines frequency-domain encoding with time-domain decoding. We identify a ``resource-scattering'' bottleneck in frequency-domain decoders under extreme computational budgets and demonstrate that replacing it with a time-domain decoder mitigates this issue, yielding improved modeling efficiency and reconstruction fidelity at comparable complexity.

  \item \textbf{Cyclic Calibration and Refinement (CCR) Training Strategy.}  
  To address the challenges arising from multi-objective optimization, we design a CCR strategy that periodically suspends adversarial losses and uses the reconstruction loss as a dynamic regularizer. This cyclic process encourages parameter exploration and re-projection across loss surfaces, systematically enhancing optimization stability and extending the performance ceiling.

  \item \textbf{Task-Driven Noise-Invariant Fine-Tuning.}  
  At extremely low bitrates (e.g., 1~kbps), the main limitation lies in bit allocation efficiency. After baseline convergence on clean speech, we introduce noisy and reverberant samples during fine-tuning while maintaining clean speech as the reconstruction target. This forces the model to learn representations that are invariant to noise and reverberation.
\end{itemize}

\begin{figure*}[t]
    \centering
    \includegraphics[width=\textwidth]{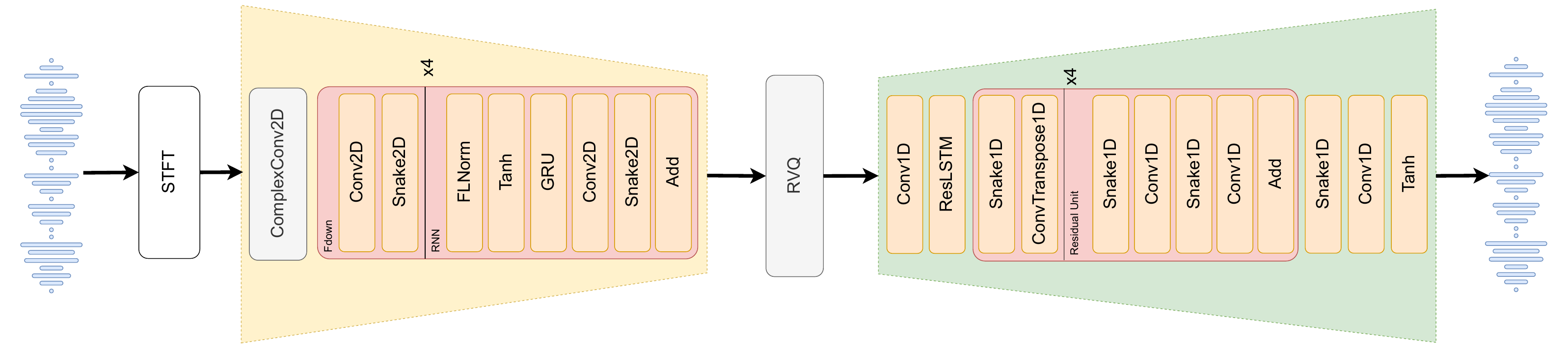}
    \caption{Overall pipeline and architectural design of PhoenixCodec.}
    \label{fig:Framework}
\end{figure*}
\section{Method}
\label{sec:method}

\subsection{Problem Formulation}
\label{ssec:problem_formulation}

The objective of this work is to address the standard speech reconstruction problem. For a codec $C(\cdot;\theta)$, the objective is to minimize the distortion $d(x, \hat{x})$ between the input signal $x$ and its reconstruction $\hat{x}$. The input $x$ may contain clean speech $s$, optionally mixed with additive noise $n$ and/or reverberation $r$. The standard optimization objective can be formulated as follows:
\begin{equation}
\mathcal{L}_{\text{standard}}(\theta) = \mathbb{E}_{x}\big[d(x, C(x;\theta))\big].
\end{equation}
Under limited computational resources, effectively optimizing this objective presents significant challenges for both model design and training strategy.

\subsection{Architecture}
\label{ssec:architecture}

To address these challenges, we begin with architectural optimization. The proposed \textbf{PhoenixCodec} (Figure~\ref{fig:Framework}) introduces a significant architectural advance compared to our previous framework, SpecTokenizer~\cite{wan2025spectokenizer}, particularly under stringent computational constraints. We observe that the primary bottleneck stems from the inherent design of the frequency-domain decoder, which severely limits the effectiveness of the frequency-to-frequency (F--F) symmetric architecture under such extreme resource constraints.

\subsubsection{Structural Bottleneck of Frequency-Grouping Decoders}
\label{sssec:capacity_diffusion}

We analyze the limitations of the frequency-domain decoder architecture adopted from our baseline, SpecTokenizer~\cite{wan2025spectokenizer}. This design employs a fixed frequency grouping strategy, partitioning the spectrum into sub-bands processed by identical neural blocks. While effective given sufficient compute budgets, under the strict constraints of this challenge (e.g., decoder $\le$ 300 MFLOPs), this uniform allocation leads to \textit{resource-scattering}, where the resources assigned to each sub-band are insufficient to effectively capture its internal spectral characteristics. Consequently, the model's representational capacity is inefficiently dispersed across the entire spectrum, resulting in inadequate modeling depth in critical resonant regions of speech and degraded synthesis fidelity.

\subsubsection{Architecture Evolution: Frequency--Time Design}
\label{sssec:ft_design}

To overcome this bottleneck, we introduce an asymmetric frequency–time architecture that preserves the frequency-domain encoder while replacing the decoder with an efficient time-domain counterpart. The proposed design provides a key advantage of computationally efficient waveform generation, enabling unified utilization of resources during synthesis. Through data-driven training, the model implicitly learns to allocate its resources toward perceptually important signal regions, thereby effectively mitigating the resource-scattering issue inherent in frequency-domain approaches.

In implementation, we adopt and refine a scalable time-domain decoder architecture inspired by BigCodec \cite{xin2024bigcodec}. To meet computational constraints, both the parameter count and the computational load are substantially reduced. In addition, a streaming adaptation is introduced to satisfy the strict real-time latency requirement of less than 30~ms for interactive scenarios.

\subsection{Training Methodology}
\label{ssec:training_method}

After establishing an efficient frequency--time architecture, the next key challenge is effective parameter optimization.

\subsubsection{Composite Optimization Objective}
\label{sssec:loss}

Our model’s optimization objective comprises four losses: the multi-resolution Mel-spectrogram reconstruction loss ($L_{\text{mel}}$) \cite{xin2024bigcodec}, the vector quantization loss ($L_{\text{vq}}$), the feature matching loss ($L_{\text{fm}}$), and the adversarial loss ($L_{\text{adv}}$). The total generator loss $L_G$ is formulated as:
\begin{equation}
L_G = L_{\text{mel}} + \lambda_{\text{vq}} L_{\text{vq}} + \lambda_{\text{fm}} L_{\text{fm}} + \lambda_{\text{adv}} L_{\text{adv}}.
\end{equation}

$L_{\text{mel}}$ and $L_{\text{vq}}$ remain active throughout all stages, while the weights $\lambda_{\text{fm}}$ and $\lambda_{\text{adv}}$ are dynamically scheduled by the training strategy. 
Our discriminator consists of a Multi-Period Discriminator (MPD) and a Multi-Resolution Discriminator (MRD) \cite{kong2020hifigangenerativeadversarialnetworks}.

\subsubsection{Cyclical Calibration and Refinement}
\label{sssec:ccr_strategy}
The CCR process consists of three sequential stages:

\noindent
\textbf{Stage I -- Pre-training.}  
During the initial stage, only the reconstruction and quantization losses are activated 
(i.e., $\lambda_{\text{fm}} = 0$, $\lambda_{\text{adv}} = 0$), enabling the model to establish a stable spectral representation.

\noindent
\textbf{Stage II -- Joint Objective Optimization.}  
When validation performance converges, all loss terms—including feature-matching and adversarial components—are re-enabled. 
This stage enhances perceptual fidelity through comprehensive optimization.

\noindent
\textbf{Stage III -- Cyclical Refinement.}  
After convergence of joint training, the model enters iterative CCR cycles. 
Distinct from the initialization phases, this stage employs a significantly reduced learning rate and fewer training steps per cycle to facilitate fine-grained optimization.
Each cycle alternates between two phases:

    \textbf{Calibration:} Disable adversarial and feature matching losses ($\lambda_{\text{fm}}=0$, $\lambda_{\text{adv}}=0$) and perform fine-tuning solely using $L_{\text{mel}}$ and $L_{\text{vq}}$ to stabilize the spectral representation.
    
    \textbf{Refinement:} Re-enable all objectives ($L_G$) to restore perceptual optimization.

The cyclic process is repeated until no further improvement is observed.
Conceptually, CCR functions as an \textbf{alternating optimization strategy}. 
The discriminator-based losses ($L_{\text{adv}}$ and $L_{\text{fm}}$) drive the model to explore high-fidelity acoustic textures and details. Conversely, the reconstruction loss ($L_{\text{mel}}$) acts as a stabilizing constraint during the calibration phase, ensuring the generated content remains consistent with the target speech.
This alternating optimization allows the model to balance perceptual quality with signal fidelity, progressively approaching its performance bound.

\subsection{Noise-Invariant Fine-Tuning}
\label{ssec:ni_ft}

While CCR provides a robust training framework, we further introduce a final fine-tuning stage aimed at maximizing perceptual performance—especially at the extreme 1~kbps bitrate.

\subsubsection{Core Hypothesis}
\label{sssec:ni_hypothesis}

We hypothesize that at extreme compression (e.g., 1~kbps), the dominant bottleneck in speech quality shifts from reconstruction capacity to \textit{bit allocation efficiency}. Any bits spent encoding non-target acoustic content such as noise or reverberation will diminish fidelity of core speech information.

\subsubsection{Baseline Training and Task Fine-Tuning}
\label{sssec:ni_training}

The process consists of two stages:

\textbf{(1) Baseline Speech Codec Training:} The model is first trained on clean speech data ($s$) to optimize the standard reconstruction target $d(s, C(s;\theta))$ until convergence.
    
\textbf{(2) Noise-Invariant Fine-Tuning:} The dataset is then extended to include noisy ($s+n$) and reverberant ($s+r$) samples, while keeping the reconstruction targets unified as clean speech $s$. For all inputs $x$, the model is fine-tuned on $d(s, C(x;\theta))$ until re-convergence.

\subsubsection{Principle and Validation}
\label{sssec:ni_validation}

This fine-tuning enforces learning of a noise- and reverberation-invariant representation at the model’s information bottleneck, improving bit allocation efficiency and yielding two benefits: 
(1) for clean speech, it guides the model to utilize more valid codes to encode core speech features, thereby enhancing fidelity; 
(2) for noisy or reverberant inputs, it improves robustness by suppressing non-speech components. 

This efficient bit allocation mechanism enables our system, in the \textit{LRAC~2025 Challenge Track~1}, to achieve first place under the most demanding 1~kbps setting in both \textit{Real-world light noise and reverb} and \textit{Intelligibility in clean speech} subjective evaluations. The effectiveness of this strategy will be quantitatively validated in the ablation study (Section~\ref{sssec:ablation_nif}).

\section{Experiments}
\label{sec:experiments}

\subsection{Experimental Setup}
\label{ssec:exp_setup}

\noindent\textbf{Dataset and Metrics.}
All training data are sourced from the official \textit{LRAC~2025} dataset~\cite{wojcicki2025lowresourceaudiocodeclrac}. During fine-tuning, we mix clean, noisy, and reverberant speech (1:1:1 ratio) with SNRs uniformly sampled between 10 dB and 30 dB.
Objective evaluations use the Versa toolkit~\cite{shi2025versaversatileevaluationtoolkit} (PESQ, UTMOS, Scoreq\_Ref). The Word Error Rate (WER) was computed on the LibriSpeech test set using Whisper~Large~V3. Subjective evaluations (MUSHRA, DMOS, DRT) were crowdsourced by the organizers.

\noindent\textbf{Implementation Details.}
The proposed system (1.48 M parameters, 699 MFLOPs) operates at 24 kHz with a frame shift of 288 samples (approx. 83 Hz). 
The encoder employs channel sizes of [32, 32, 32, 128, 335]. 
The RVQ module consists of 6 codebooks, each with a size of 4096 and a dimension of 8.
The decoder features channels of [117, 58, 29, 14, 7] with upsampling rates of [3, 4, 4, 4]. 
To ensure strict latency compliance (29.83 ms total), we implement an 11th-order IIR low-pass filter for efficient resampling instead of standard FIR approaches.
Discriminator architectures and loss functions follow the configurations in BigCodec~\cite{xin2024bigcodec}.
Regarding optimization, we use the Adam optimizer throughout all stages. The initial learning rate is set to $8 \times 10^{-4}$ for Stage I and $1 \times 10^{-4}$ for Stages II and III, decaying to $1 \times 10^{-5}$. Stages I and II are each trained for approximately 200k steps, while in Stage III, each phase lasts roughly 10k steps. The cycle repeats until validation metrics and loss stabilize.

\noindent\textbf{Baselines.}
We compare against two symmetric baselines under identical constraints:
(1) \textbf{F--F}: A compliant frequency-domain architecture based on SpecTokenizer~\cite{wan2025spectokenizer};
(2) \textbf{T--T}: A compliant time-domain architecture based on BigCodec~\cite{xin2024bigcodec}.
Further details are available at the \textcolor{magenta}{\href{https://ggiggit.github.io/phoenixcodec.github.io/}{project page}}.

\subsection{Main Results}
\label{ssec:main_results}
\begin{table}[!t]
\centering
\caption{Evaluation results in \textit{LRAC 2025 Challenge Track 1}.}
\vspace{4pt}
\label{tab:main}
\resizebox{\columnwidth}{!}{%
\begin{tabular}{
    l!{\vrule width 0.8pt}
    c c!{\vrule width 0.8pt}
    c c!{\vrule width 0.8pt}
    c c!{\vrule width 0.8pt}
    c
}
\toprule
\multicolumn{1}{l!{\vrule width 0.8pt}}{} 
& \multicolumn{2}{c!{\vrule width 0.8pt}}{Clean} 
& \multicolumn{2}{c!{\vrule width 0.8pt}}{Noisy} 
& \multicolumn{2}{c!{\vrule width 0.8pt}}{Multi-talkers}
& \multicolumn{1}{c}{Intelligibility} \\
  & \multicolumn{2}{c!{\vrule width 0.8pt}}{MUSHRA}
  & \multicolumn{2}{c!{\vrule width 0.8pt}}{DMOS}
  & \multicolumn{2}{c!{\vrule width 0.8pt}}{DMOS}
  & DRT score \\
\cmidrule(lr){1-1}
\cmidrule(lr){2-3}
\cmidrule(lr){4-5}
\cmidrule(lr){6-7}
\cmidrule(lr){8-8}
Bitrate
& 1\,kbps & 6\,kbps
& 1\,kbps & 6\,kbps
& 1\,kbps & 6\,kbps
& 1\,kbps \\
\midrule
Baseline
  & 17.92 & 74.28
  & 1.31 & 3.35
  & 1.26 & 2.20
  & 75.90 \\
1st place
  & 62.65 & 81.75
  & 3.02 & 4.44
  & 2.82 & 4.35
  & 85.43 \\
\textbf{Proposed}
  & 60.90 & 80.69
  & \textbf{3.40} & 4.16
  & 2.08 & 2.98
  & \textbf{85.57} \\
\bottomrule
\end{tabular}
}
\vspace{2pt}
\end{table}  

Table~\ref{tab:main} summarizes the evaluation results in the \textit{LRAC~2025 Challenge Track~1} across clean, noisy, and multi‑talker conditions, along with intelligibility assessed by the Diagnostic Rhyme Test (DRT). The proposed system delivers consistently competitive performance, especially in the extreme low‑bitrate scenario. Notably, it achieves the highest DMOS score under noisy conditions at 1~kbps, as well as the highest DRT score, outperforming both the baseline and the first‑place reference systems. These results confirm the effectiveness of the proposed architecture and its robustness in perceptually challenging environments.
More details about the leaderboard and other systems are available at the official results page~\footnote{\url{https://crowdsourcing.cisco.com/lrac-challenge/2025/results\#track-1--transparency-codecs}}.

\subsection{Ablation Studies}
\label{ssec:ablation}

To assess the effectiveness of the proposed F‑T architecture, CCR strategy, and noise‑invariant fine‑tuning, we conducted ablation studies using 40\% of the clean speech data for Sections~\ref{sssec:ablation_ft} and~\ref{sssec:ablation_ccr}, and 40\% of the entire dataset for Section~\ref{sssec:ablation_nif} to reduce computational cost.  
Metrics with the suffixes~$c$~and~$n$~denote evaluations on clean and noisy speech, respectively. For clarity, Tables~\ref{tab:ab1} and~\ref{tab:ab3} present only the results of the first stage of CCR, as the subsequent stages follow similar performance trends.

\begin{table}[t]
\centering
\caption{Performance comparison of different architectures.}
\label{tab:ab1}
\setlength{\tabcolsep}{2.5pt}
\begin{adjustbox}{max width=\linewidth}
\begin{tabular}{lcccccccc}
\toprule
Bitrate & Method & WER & PESQ$_c$ & UTMOS$_c$ & Scoreq$_c$ & PESQ$_n$ & UTMOS$_n$ & Scoreq$_n$ \\
\midrule
& GT & 2.89 & 4.64 & 4.03 & 0.00 & 2.24 & 3.30 & 0.63 \\ 
\midrule
\multirow{3}{*}{1 kbps}
 & F‑F & \textbf{7.41} & 1.73 & 2.32 & 0.68 & 1.42 & 1.96 & 0.94 \\
 & T‑T & 10.87 & 1.92 & 2.94 & 0.50 & 1.49 & 2.55 & 0.83 \\
 & F‑T & 7.47 & \textbf{1.98} & \textbf{2.99} & \textbf{0.49} & \textbf{1.53} & \textbf{2.61} & \textbf{0.81} \\
\midrule
\multirow{3}{*}{6 kbps}
 & F‑F & \textbf{3.14} & 2.39 & 2.80 & 0.50 & 1.71 & 2.45 & 0.83 \\
 & T‑T & 3.50 & 2.79 & 3.41 & 0.35 & 1.88 & 2.97 & 0.78 \\
 & F‑T & 3.24 & \textbf{2.80} & \textbf{3.45} & \textbf{0.33} & \textbf{1.90} & \textbf{3.04} & \textbf{0.70} \\
\bottomrule
\end{tabular}
\end{adjustbox}
\vspace{2pt}
\end{table}
\begin{table}[t]
\centering
\caption{Performance comparison across training stages. CCR stages are fine-tuned sequentially from previous checkpoints, while \textit{Joint-opt} is trained from scratch. \textit{CCR}$_{s1}$ and \textit{CCR}$_{s2}$ denote states after Stage I and II; \textit{CCR}$_{s3\text{-}c}$ and \textit{CCR}$_{s3\text{-}r}$ represent the Calibration and Refinement phases of Stage III.}
\label{tab:ab2}
\setlength{\tabcolsep}{2.8pt}
\begin{adjustbox}{max width=\linewidth}
\begin{tabular}{lcccccccc}
\toprule
Bitrate & Method & WER & PESQ$_c$ & UTMOS$_c$ & Scoreq$_c$ & PESQ$_n$ & UTMOS$_n$ & Scoreq$_n$ \\
\midrule
& GT & 2.89 & 4.64 & 4.03 & 0.00 & 2.24 & 3.30 & 0.63 \\ 
\midrule
\multirow{5}{*}{1 kbps}
 & Joint‑opt  & 16.49 & 1.72 & 2.87 & 0.55 & 1.37 & 2.29 & 0.93 \\
 & CCR$_{\text{s1}}$ & \textbf{7.47} & 1.98 & 2.99 & 0.49 & 1.53 & 2.61 & 0.81 \\
 & CCR$_{\text{s2}}$ & 8.86 & 1.94 & 3.29 & 0.40 & 1.50 & 2.67 & 0.81 \\
 & CCR$_{\text{s3-c}}$ & 8.03 & \textbf{2.13} & \textbf{3.44} & 0.38 & \textbf{1.60} & \textbf{2.89} & \textbf{0.76} \\
 & CCR$_{\text{s3-r}}$ & 8.74 & 2.02 & 3.38 & \textbf{0.37} & 1.51 & 2.69 & 0.79 \\
\midrule
\multirow{5}{*}{6 kbps}
 & Joint‑opt  & 4.25 & 2.55 & 3.66 & 0.26 & 1.80 & 3.08 & 0.72 \\
 & CCR$_{\text{s1}}$ & 3.24 & 2.80 & 3.45 & 0.33 & 1.90 & 3.04 & 0.70 \\
 & CCR$_{\text{s2}}$ & 3.62 & 2.92 & 3.76 & 0.22 & 1.89 & 3.14 & 0.70 \\
 & CCR$_{\text{s3-c}}$ & \textbf{3.22} & \textbf{3.09} & \textbf{3.82} & 0.22 & \textbf{1.96} & \textbf{3.24} & \textbf{0.67} \\
 & CCR$_{\text{s3-r}}$ & 3.27 & 2.98 & 3.76 & \textbf{0.21} & 1.90 & 3.18 & 0.69 \\
\bottomrule
\end{tabular}
\end{adjustbox}
\vspace{2pt}
\end{table}
\begin{table}[t]
\centering
\caption{Performance impact of noise-invariant fine-tuning (NIFT).}
\label{tab:ab3}
\setlength{\tabcolsep}{2.5pt}
\begin{adjustbox}{max width=\linewidth}
\begin{tabular}{lcccccccc}
\toprule
Bitrate & Method & WER & PESQ$_c$ & UTMOS$_c$ & Scoreq$_c$ & PESQ$_n$ & UTMOS$_n$ & Scoreq$_n$ \\
\midrule
& GT & 2.89 & 4.64 & 4.03 & 0.00 & 2.24 & 3.30 & 0.63 \\ 
\midrule
\multirow{2}{*}{1 kbps}
 & w/o NIFT & \textbf{5.89} & \textbf{1.80} & 2.13 & 0.73 & 1.46 & 1.76 & 0.98 \\
 & w/  NIFT & 8.52 & 1.78 & \textbf{2.39} & \textbf{0.67} & \textbf{1.60} & \textbf{2.29} & \textbf{0.73} \\
\midrule
\multirow{2}{*}{6 kbps}
 & w/o NIFT & \textbf{3.34} & \textbf{2.44} & 2.66 & 0.53 & 1.70 & 2.32 & 0.85 \\
 & w/  NIFT & 3.35 & 2.42 & \textbf{2.89} & \textbf{0.49} & \textbf{2.06} & \textbf{2.88} & \textbf{0.55} \\
\bottomrule
\end{tabular}
\end{adjustbox}
\vspace{2pt}
\end{table}
\subsubsection{Effectiveness of the F--T Architecture}
\label{sssec:ablation_ft}

As presented in Table~\ref{tab:ab1}, the F-T architecture demonstrates the most effective balance between perceptual quality and semantic fidelity under the challenge constraints.
We observe distinct trade-offs among the architectures: the F-F model maintains a competitive WER (7.41\%), highlighting the frequency-domain encoder's strength in capturing semantic content. However, its perceptual metrics are lower due to the structural limitations of the decoder discussed in Section~\ref{sssec:capacity_diffusion}.
Conversely, the T-T model offers improved perceptual quality but suffers from a significantly higher WER (10.87\%), suggesting that pure time-domain encoding may struggle to preserve semantic information at extremely low bitrates.

The proposed F-T design leverages the best of both worlds: it utilizes frequency-domain encoding to preserve semantic integrity (achieving WER comparable to F-F) while employing a time-domain decoder for efficient high-fidelity waveform generation. This results in an optimal trade-off, outperforming symmetric baselines in overall utility for this specific low-resource scenario.

\subsubsection{Effectiveness of the CCR Training Strategy}
\label{sssec:ablation_ccr}

From the results in Table~\ref{tab:ab2}, several key observations can be made:
(1) \textit{CCR}$_{s2}$ outperforms the \textit{Joint-opt} baseline. This indicates that direct multi-objective optimization tends to hinder stable convergence, whereas the proposed stage-wise optimization strategy provides a more efficient learning trajectory under identical resource constraints.
(2) The CCR mechanism further enhances performance beyond the convergence limit of Stage II. Compared with stage-wise configurations (\textit{CCR}$_{s1}$ vs.\ \textit{CCR}$_{s3\text{-}c}$, \textit{CCR}$_{s2}$ vs.\ \textit{CCR}$_{s3\text{-}r}$), the cyclic approach demonstrates greater stability and generalization. This validates the effectiveness of alternating between reconstruction constraints and adversarial refinement, consistent with the design motivation discussed in Section~\ref{sssec:ccr_strategy}.

Although \textit{CCR}$_{s3\text{-}c}$ achieves the best objective scores across most metrics, its outputs sound slightly mechanical due to the exclusive use of $L_\text{mel}$; re-enabling all objectives in the \textit{CCR}$_{s3\text{-}r}$ refinement stage yields noticeably more natural audio. Overall, CCR's alternating optimization strategy enables more stable training and better potential utilization under constrained resources.

\subsubsection{Effectiveness of Noise-Invariant Fine-Tuning}
\label{sssec:ablation_nif}

Table~\ref{tab:ab3} shows that at 1~kbps and 6~kbps, perceptual quality under noisy conditions improves markedly, while clean‑speech performance remains stable or slightly higher.
These results suggest that NIFT effectively enhances noise robustness and bit‑allocation efficiency without sacrificing reconstruction fidelity.
While the WER at~1~kbps~increases due to a trade‑off between acoustic detail and semantic consistency, Table~\ref{tab:main}~demonstrates that its effect on intelligibility remains limited, with the proposed system achieving top scores at~1~kbps.

\section{Conclusion}
\label{sec:conclusion}

To address neural speech coding under extremely low-resource conditions, we propose PhoenixCodec, a unified framework that integrates architectural, training, and fine-tuning innovations.
An efficient asymmetric frequency–time design effectively utilizes computational capacity at the decoder to overcome the ``resource-scattering'' bottleneck in conventional frequency-domain methods.
The CCR strategy enables stable optimization and improves overall performance, while noise-invariant fine-tuning ensures robust reconstruction under severe compression.
Ablation results confirm the independent contributions of each component, demonstrating that PhoenixCodec achieves a well-balanced trade-off between reconstruction quality and computational efficiency.

\bibliographystyle{IEEEbib}
\bibliography{paper}

\end{document}